# Enumeration Order Equivalency

**Ali Akbar Safilian**
(AmirKabir University of Technology, Tehran, Iran
ali_safilian@aut.ac.ir)

**Farzad Didehvar**
(AmirKabir University of Technology, Tehran, Iran
didehvar@aut.ac.ir)

Let $A \subseteq N$ be an infinite recursively enumerable set. There are some total computable functions $h: N \to A$ connected with a Turing machine such that $A = \{h(1), h(2), \ldots\}$.

**Definition 2.1 (Listing)** A *listing* of an infinite r.e. set $A \subseteq N$ is a bijective computable function $f: N \to A$.

As mentioned above we can connect every Turing machine with a listing uniquely. Namely, each listing shows the enumeration order of the elements enumerated by the related Turing machine. In the following, we define a reduction on listings and sets based on enumeration orders.

**Definition 2.2 (Enumeration Order Reducibility on Listings and sets)**
1. For listings $f, g: N \to N$ we say $f$ is "Enumeration order reducible" to $g$ and write $f \leq_{eo} g$, if and only if, $\forall i < j, (f(i) > f(j) \Rightarrow g(i) > g(j))$.
2. For r.e. sets $A, B \subseteq N$, we say $A$ is "Enumeration order reducible" to $B$ and write $A \leq_{eo} B$, if and only if, for any listing $g$ of $B$ there exist some computable function $F_{AB}$ (from the listings of $B$ to the listings of $A$) such that $F_{AB}(g) \leq_{eo} g$.
3. For two listings $f, g$ (r.e. sets $A, B$), we say $f$ is "Enumeration order equivalent" to $g$ and write $f \equiv_{eo} g$ ($A \equiv_{eo} B$), if and only if, both $f \leq_{eo} g$ and $g \leq_{eo} f$ ($A \leq_{eo} B$ and $B \leq_{eo} A$)

Let $A$ be an r.e. subset of $N$. $[A]_{eo} = \{B \subseteq N | A \equiv_{eo} B\}$ denotes the *Enumeration order equivalence class* of $A$. Also we call the equivalence classes of enumeration order equivalent sets "Enumeration order degree" and write $deg(A)$ for an r.e. set $A$. The relation $\leq_{eo}$ is preorder, because it has transitive and reflexive properties but it is not a partial order because $A \leq_{eo} B$ and $B \leq_{eo} A$ does not necessarily imply $A = B$.

**Lemma 2.3** Any decidable set is enumeration order reducible to every r.e. set.
**Proof:** It is a clear fact that there exists a listing $f$ for any decidable set $A$ such that enumerates the elements of $A$ in the ascendant usual order. Since such listing is enumeration order reducible to every listing, all decidable sets are enumeration order reducible to every r.e. set. □

Depending on Definition 2.2, it is clear that if $A <_{eo} B$ then $A$ is $B -$ recursive, so the following proposition is valid.

**Proposition 2.4** If $A <_{eo} B$, then $A <_T B$. □

From the above proposition we can conclude the following propositions:

**Proposition 2.5** The enumeration order degree of decidable sets is the least one. The enumeration order degree of decidable sets is denoted by $[\emptyset]_{eo}$. Also, the enumeration order degree $[K]_{ue}$ is the maximal one. □

**Proposition 2.7** There are infinite chains on this relation. □

The upcoming theorem is the one of the main theorems in this paper which influences the study of the related equivalency. But before of that we need to depict the following lemma.

**Lemma 2.8** Consider two listings $f, g$ of two r.e. sets $A, B$ respectively such that $f \leq_{eo} g$. Assume that we show the sets $A, B$ with two sequences of natural numbers $\{a_i\}_{i \in N}$ and $\{b_i\}_{i \in N}$ respectively such that for all $i \in N$: $a_i < a_{i+1}$ and $b_i < b_{i+1}$, then:
1) $f^{-1}(a_1) \leq g^{-1}(b_1)$ and
2) For all $i > 1$, if for all $j < i$: $f^{-1}(a_j) = g^{-1}(b_j)$ then $f^{-1}(a_i) \leq g^{-1}(b_i)$.

**Proof:**
First we want to prove that $f^{-1}(a_1) \leq g^{-1}(b_1)$. For the sake of a contradiction, assume that $f^{-1}(a_1) > g^{-1}(b_1)$. Since $f(g^{-1}(b_1)) > f(f^{-1}(a_1))$ and $f \leq_{eo} g$, $g(g^{-1}(b_1)) > g(f^{-1}(a_1))$. But this is a contradiction, because $b_1$ is the least element of $B$.

Consider a number $i > 1$. Assume that for all $j < i$: $f^{-1}(a_j) = g^{-1}(b_j)$. We want to prove $f^{-1}(a_i) \leq g^{-1}(b_i)$. For the sake of a contradiction, assume that $f^{-1}(a_i) > g^{-1}(b_i)$. It is clear that $f(g^{-1}(b_i)) \notin \{a_1, a_2, \dots, a_{i-1}\}$ and $g(f^{-1}(a_i)) \notin \{b_1, b_2, \dots, b_{i-1}\}$. Therefore, $f(g^{-1}(b_i)) > f(f^{-1}(a_i))$. Since $f \leq_{eo} g$, $g(g^{-1}(b_i)) > g(f^{-1}(a_i))$. This is a contradiction. □

**Theorem 2.9** Consider two enumeration order equivalent sets $A$ and $B$. There exist two listings $f, g$ of $A, B$ respectively such that $f \equiv_{eo} g$

**Proof:**
Consider a listing $f$ of $A$. According to Definition 2.2, we can deduce that there are two sequences $\{f_i\}_{i \in N}$ and $\{g_i\}_{i \in N}$ of listings of $A, B$ respectively such that $f_1 = f$ and $\dots \leq_{eo} g_2 \leq_{eo} f_2 \leq_{eo} g_1 \leq_{eo} f_1$. Therefore, the following statement is valid:
$\dots \leq_{eo} f_3 \leq_{eo} f_2 \leq_{eo} f_1$ [1]

According to Definition 2.2 (Definition of Enumeration Order Reducibility on r.e. sets) the chain [1] is a computable chain.

Based on Lemma 2.8 and this fact that [1] is an infinite chain, we can deduce that there exists $n \in N$ such that for all $i \geq n$: $f_i^{-1}(a_1) = f_{i+1}^{-1}(a_1)$.

Assume that there exist $n, k \in N$ such that for all $j \leq k$ and $i \geq n$: $f_i^{-1}(a_j) = f_{i+1}^{-1}(a_j)$. Now consider the chain $\dots \leq_{eo} f_{n+2} \leq_{eo} f_{n+1} \leq_{eo} f_n$ [2]. Once again based on Lemma 2.8 and this fact that [2] is an infinite chain, we can gather that there exists $m \in N$ such that for all $i \geq m + n$: $f_i^{-1}(a_{k+1}) = f_{i+1}^{-1}(a_{k+1})$. [*]

Based on [*], there are $i, j \in N$ such that $i < j$ and $f_i = f_j$ in the chain [1]. As mentioned above this chain is computable, so $i$ and $j$ can be obtained computably. Since enumeration order reducibility is transitive, $f_i \leq_{eo} g_i \leq_{eo} f_i$, $f_i \equiv_{eo} g_i$. □

**Theorem 2.10** For r.e. sets $A$ and $B$ the following statements are equivalent:
1. $A \equiv_{eo} B$,
2. For every listing $f$ of $A$ there is a listing $g$ of $B$ such that $f \equiv_{eo} g$.

**Proof:**
**1⇒2:** Let $A \equiv_{eo} B$. According to Theorem 2.9, there exist listings $h'$ of $A$ and $g'$ of $B$ such that $h' \equiv_{eo} g'$. Assume that $h$ is a listing of $A$. We want to define a listing $g$ of $B$ such that $h \equiv_{eo} g$. Consider $g = g'oh'^{-1}oh$. It is evident that $g$ is a listing of $B$. We claim that $h \sim g$. Assume that for $, j \in N$, $h(i) < h(j)$ and $h(i) = a_1 \in A$, $h(j) = a_2 \in A$. Since listings are surjective functions, there exist $m, k \in N$ such that $h'(k) = a_1$ and $h'(m) = a_2$.

$$h(i) < h(j) \Rightarrow h'(k) < h'(m) \Rightarrow g'(k) < g'(m) \Rightarrow g'oh'^{-1}(a_1) < g'oh'^{-1}(a_2)$$
$$\Rightarrow g'oh'^{-1}h(i) < g'oh'^{-1}h(j) \Rightarrow g(i) < g(j)$$

Therefore, $h \equiv_{eo} g$.
**2⇒1:** it is evident. □

We illustrate the above concepts by the following example.
**Example 2.11**
1) Two sets $A = \{2i : i \in N\}$ and $B = N - \{1\}$ are enumeration order equivalent.

**Proof:** Consider two listings $h$ of $A$ and $g$ of $B$ such that for all $i \in N$, $h(i) = 2i$ and $g(i) = i + 1$. It is clear that for all $i, j \in N$, $h(i) < h(j) \Leftrightarrow g(i) < g(j)$ □

2) Two sets $N$ and $K$ are not enumeration order equivalent.

**Proof:** For the sake of a contradiction, assume that these two sets are enumeration order equivalent. The identity function $id: N \to N$ is a listing of $N$. There exists a listing $g$ of $K$ such that $h \equiv_{eo} g$. Therefore, for all $i, j \in N$, $i < j \Leftrightarrow g(i) < g(j)$. But this cannot be valid, because $k$ is not a decidable set. This causes a contradiction.□

In the continuation of this section, we want to explore some relationships between the enumeration order equivalency on sets and both one-one reducibility & Turing-reducibility.

**Lemma 2.12** If two sets belong to same one-one reducibility equivalence class, then they do not belong necessarily to same degree of enumeration order.
**Proof:** Consider an r.e. non-decidable set $A$. There are two cases.
Case 1: $1 \in A$,
Case 2: $1 \notin A$.
For simplicity, we assume case 2. Two sets $A$ and $B = A \cup \{1\}$ belong to same one-one reducibility equivalence class. Let $A = \{a_1, a_2, \ldots, a_n, \ldots\}$ in which $a_1 < a_2 < \cdots < a_n < \cdots$ and $f: N \to B$ is a listing of $B$ in which $f(1) = 1$. For the sake of a contradiction, assume that two sets $A$ and $B$ are enumeration order equivalent. Therefore, there exists a listing $g$ of $A$ such that $f \equiv_{eo} g$. Since $f(1)$ is the minimum element of $B$, $g(1)$ should be $a_1$. Assume that $f(2) = a_n$. Since two listings $f, g$ are enumeration order equivalent, $g(2)$ should be $a_{n-1}$(It is evident that for all $i \in N$, the cardinality of two sets $\{j|\ f(j) < f(i)\}$ and $\{j|\ g(j) < g(i)\}$ are same.). It is evident that we can compute a number $s_1$ such that $f(s_1) = a_{n-1}$. Again, $g(s_1)$ should be $a_{n-2}$. We can keep this way for $a_{n-1}$ and find the number $s_2$ such that $g(s_2) = a_{n-2}$. And etc…

The above issue, shows that for an element $a_i \in A$, we can compute the lesser elements than $a_i$ that are belong to $A$. This shows that $A$ is a decidable set and this causes a contradiction.

For case 1, we can consider $A' = A - \{1\}$ and deal with it in a similar way instead of $A$. □

In Example 2.11 we showed that two sets $A = \{2i: i \in N \}$ and $B = N - \{1\}$ are enumeration order equivalent. It is clear that these two sets do not belong to same one-one reducibility equivalence class, because the cardinality of their complement sets are not same. Therefore, we can support the following proposition.

**Proposition 2.13**   If two sets belong to same enumeration order degree, then they do not belong necessarily to same one-one reducibility equivalence class.□

According to proposition 2.4, if two r.e. sets $A$ and $B$ are enumeration order equivalent then they belong to same Turing-reducibility equivalence class.

Since if $A \equiv_{1-1} B$, then $A \equiv_T B$, we can confirm the similar result of Proposition 2.13 for Turing degrees instead of one-one degrees.

Consider an r.e. non-decidable set $A$. This question is very important: How many enumeration order equivalence classes are located in $[A]_T$.

**Theorem 2.14**   There is infinite number of enumeration order equivalence classes which are located in any Turing equivalence class.

**Proof:** Let $A$ be an r.e. non-decidable set. According to Lemma 2.12, two sets $A - \{1\}$ and $A \cup \{1\}$ are not enumeration order equivelnt. In a similar way for any $n \in N$ every two sets of the following sets are not uniform.

1) $A - \{1, ..., n\}$
2) $A - \{2, ..., n\} \cup \{1\}$
3) $A - \{3, ..., n\} \cup \{1,2\}$
4) ….
    .
    .
    .

$n + 1)$ $A \cup \{1,2, ..., n\}$

It is clear that each introduced set above is a member of $[A]_T$. Therefore, there is infinite number of uniformity equivalence classes such that they are subsets of the equivalence class $[A]_T$. □

**Acknowledgements**

The authors wish to thank a lot Professor Klaus Weihrauch for his valuable sympathy and precise hints to prepare this paper. We benefited his very valuable comments and used them to correct our faults. Also the current format of this paper is beholden to him.

## References

[1] R. Friedberg and H. Rogers. Reducibility and Completness for sets of intergers. *Z. Math. LogiK Grundlag. Math.,* 5:117-125, 1959.

[2] H.Rogers: Theory of Recursive Functions and Effective Computability. Mc-Graw Hill, New York, 1967.

[3] J. Case. Enumeration reducibiity and partial degrees. *Arch. Math. Logic,* 2 (4): 419-439, 1970/1971.

[4] M. Ziegler. Algebraisch abgeschlossene Gruppen. In S. I. Adian, W. W. Boone, and G. Higman, editors, *word problems, II* ( *Conf. On Decision Problems in Algebra, Oxford, 1978*), volume 95 of *Stud. Logic foundations Math*., pages 449-576. North-Holland, Amsterdam, 1980.

[5] S. Cooper. Enumeration reducibility, nondeterministic computations and relative computability of partial functions. In *Recursion Theory Week* ( *Oberwolfach, 1989* ), volum 1432 of *Lecture Notes in Math.,* pages 57-110. Springer, Berlin, 1990.

[6] M. Sipser: Introduction to the Theory of Computation. PWS Publishing company. 1997

[7] S. Cooper. Computability Theory. Chapman & Hall, 2004.

[8] J. Miller. Degrees of unsolvability of continuous finctions. *J. Symbolic Logic*, 69(2): 555-584, 2004.